\documentclass[twocolumn,showpacs,amsmath,amssymb,floatfix,prl,superscriptaddress]{revtex4-2}
\usepackage[latin1]{inputenc} 
\usepackage{setspace}
\usepackage{graphicx}
\usepackage{graphics}
\usepackage{subfigure}
\usepackage{dcolumn}
\usepackage{bm}
\usepackage{color}
\usepackage{SIunits}
\usepackage[normalem]{ulem}
\usepackage{url}
\usepackage[left, modulo, pagewise]{lineno}

\bibpunct{[}{]}{,}{n}{,}{,}

\newcommand{\LOA}{Laboratoire d'Optique Appliqu\'ee, Institut Polytechnique de Paris, ENSTA-Paris, Ecole Polytechnique, CNRS, 91120 Palaiseau, France}
\newcommand{\CEA}{LIDYL, CEA, CNRS, Universit\'e Paris-Saclay, CEA Saclay, 91191 Gif-sur-Yvette, France}
\newcommand{\DAM}{CEA, DAM, DIF, F-91297 Arpajon, France}

\begin{document}

\title{Sub-laser-cycle control of relativistic plasma mirrors}

\author{L. Chopineau}\thanks{These authors contributed equally}\affiliation{\CEA}\affiliation{\DAM}
\author{G. Blaclard}\thanks{These authors contributed equally} \affiliation{\CEA}
\author{A. Denoeud}\affiliation{\CEA}\affiliation{\DAM}
\author{H. Vincenti}\affiliation{\CEA}
\author{F. Qu\'er\'e}\affiliation{\CEA}
\author{S. Haessler}\email{stefan.haessler@ensta-paris.fr}\affiliation{\LOA}

\date{\today}

\begin{abstract}
We present measurements of high-order harmonics and relativistic electrons emitted into the vacuum from a plasma mirror driven by temporally-shaped ultra-intense laser waveforms, produced by collinearly combining the main laser field with its second harmonic. We experimentally show how these observables are influenced by the phase delay between these two frequencies at the attosecond timescale, and relate these observations to the underlying physics through an advanced analysis of 1D/2D Particle-In-Cell simulations. These results demonstrate that sub-cycle shaping of the driving laser field provides fine control on the properties of the relativistic electron bunches responsible for harmonic and particle emission from plasma mirrors.
\end{abstract}

\maketitle


Plasma mirrors are overdense plasmas created at the surface of laser-ionized optically flat solid targets. They are versatile optical devices for the manipulation of ultra-intense femtosecond (fs) laser beams ($I_L \gtrsim 10^{16}$ W/cm$^2$). They are also considered as a promising path for the generation of intense attosecond light pulses through high-order harmonic generation (HHG)~\cite{tsakiris2006route,dromey2006high, thaury2007plasma, teubner2009high,chopineau2021spatio}, as well as for laser-driven particle acceleration \cite{macchi2013ion, thevenet2016vacuum, tian2012electron,zaim_few-cycle_2019}. From a fundamental standpoint, plasma mirrors represent ideally simple testbeds for ultrahigh-intensity laser-plasma interaction physics because the dynamics is confined to a thin layer at the target surface, where the plasma particles are directly exposed to the ultraintense laser field, without any prior alteration of this field~\cite{kahaly2013direct,bocoum2015spatial}.

Collective electronic motion at this plasma-vacuum interface on a sub-femtosecond timescale plays a key role in this physics. When a p-polarized laser is focused on a solid target at intensities $I_L \gtrsim 10^{18}$ W/cm$^2$, the laser-driven electron motion becomes relativistic and can be described by a push--pull process~\cite{gonoskov_ultrarelativistic_2011,thevenet2016physics,gonoskov_theory_2018}, repeating once per driving laser period. The incident laser field first pushes electrons into the plasma, piling up a dense electron bunch and creating a restoring internal plasma field. As the laser field changes sign, the combined plasma and laser fields accelerate the electron bunch to a relativistic velocity towards the vacuum. This can induce drastic temporal modulations to the reflected laser wave, which sensitively depend on the electron bunch properties and dynamics~\cite{lichters_short-pulse_1996,gordienko_relativistic_2004,thaury2010high,an2010enhanced, mikhailova_isolated_2012,edwards_x-ray_2020}:  key physical parameters are the bunch charge, velocity, and spatial extent. In the spectral domain, these periodic temporal modulations result in HHG. We will refer to this process as \textit{Relativistic Oscillating Mirror} (ROM) in the following. 

While most of the electrons then get pushed back into the plasma, a fraction of them are expelled into the vacuum at relativistic velocities. Their tight temporal locking with the laser field then lets them get further accelerated in the reflected laser field through \textit{Vacuum Laser Acceleration} (VLA)~\cite{thevenet2016vacuum}. This suggests a tight correlation of HHG and fast-electron emissions in the relativistic regime, which has indeed been observed experimentally~\cite{chopineau2019identification}.

While the understanding of this relativistic laser-plasma interaction physics has greatly advanced over the last two decades, the means of its experimental control have remained rather limited. To date, the main control knob is the scale length $L_\mathrm{g}$ of the plasma density gradient at plasma-vacuum interface, $n(x)\propto\exp[-x/L_\mathrm{g}(\tau)]$, which can be varied by adjusting the delay $\tau$ between a weaker ionizing prepulse and the main ultra-intense driving pulse~\cite{kahaly2013direct,bocoum2015spatial}. On the relevant sub-femtosecond time scale, this is of course a static parameter. An adequately fast dynamic control could be achieved with a temporally tailored driving waveform. This concept has already had great success with strong-field dynamics driven on the single-atom level at much lower intensities $\sim 10^{14}$ W/cm$^2$~\cite{bartels2000shaped, dudovich2006measuring, haessler2014optimization, jin2014waveforms, wei2013selective}. Its potential for relativistic plasma mirrors has recently been shown in numerical studies~\cite{edwards2014enhanced, edwards2016waveform} that predicted significant enhancements of the HHG efficiency, shortly after confirmed in a first experiment~\cite{yeung2017experimental}. However, no other observable than the angle-integrated HHG spectrum has been studied, leaving mostly unexplored the detailed attosecond control afforded by the additional degree of freedom of the driving optical cycle shape. 


In this Letter, we present experiments where high-order harmonics as well as relativistic electrons are generated through the interaction of a plasma mirror with ultra-intense temporally shaped optical cycles, generated by combining the fundamental laser field ($\omega_L$) with its second harmonic ($2\omega_L$). By adjusting their relative phase, we can dynamically control the collective plasma electron motion on the sub-femtosecond time scale. We simultaneously measure the angularly and spectrally resolved high-order harmonics generated around the specular direction, as well as the high-energy electron beam emitted into the vacuum. Finally, we relate these observations to the underlying physics through an advanced analysis of 1D/2D Particle-In-Cell (PIC) simulations.

The experiments are carried out on the 100 TW-class Ti:sapphire laser UHI100 (LIDYL, CEA Saclay), delivering 25-fs pulses at central wavelength $\lambda_L = 800\:$nm with a temporal contrast $\gtrsim 10^{13}$ on a $\gtrsim 100$ ps timescale~\cite{levy2007double}.  
An aperture mask transmits a 33-mm-diameter top-hat main beam (limited by the available size of the calcite crystal, see below) as well the weak prepulse, used to control the gradient scale length $L_g$~\cite{kahaly2013direct}. The value of $L_g$ was measured using spatial domain interferometry~\cite{bocoum2015spatial}. 

\begin{figure}
	\centering
		\includegraphics[width=1\columnwidth]{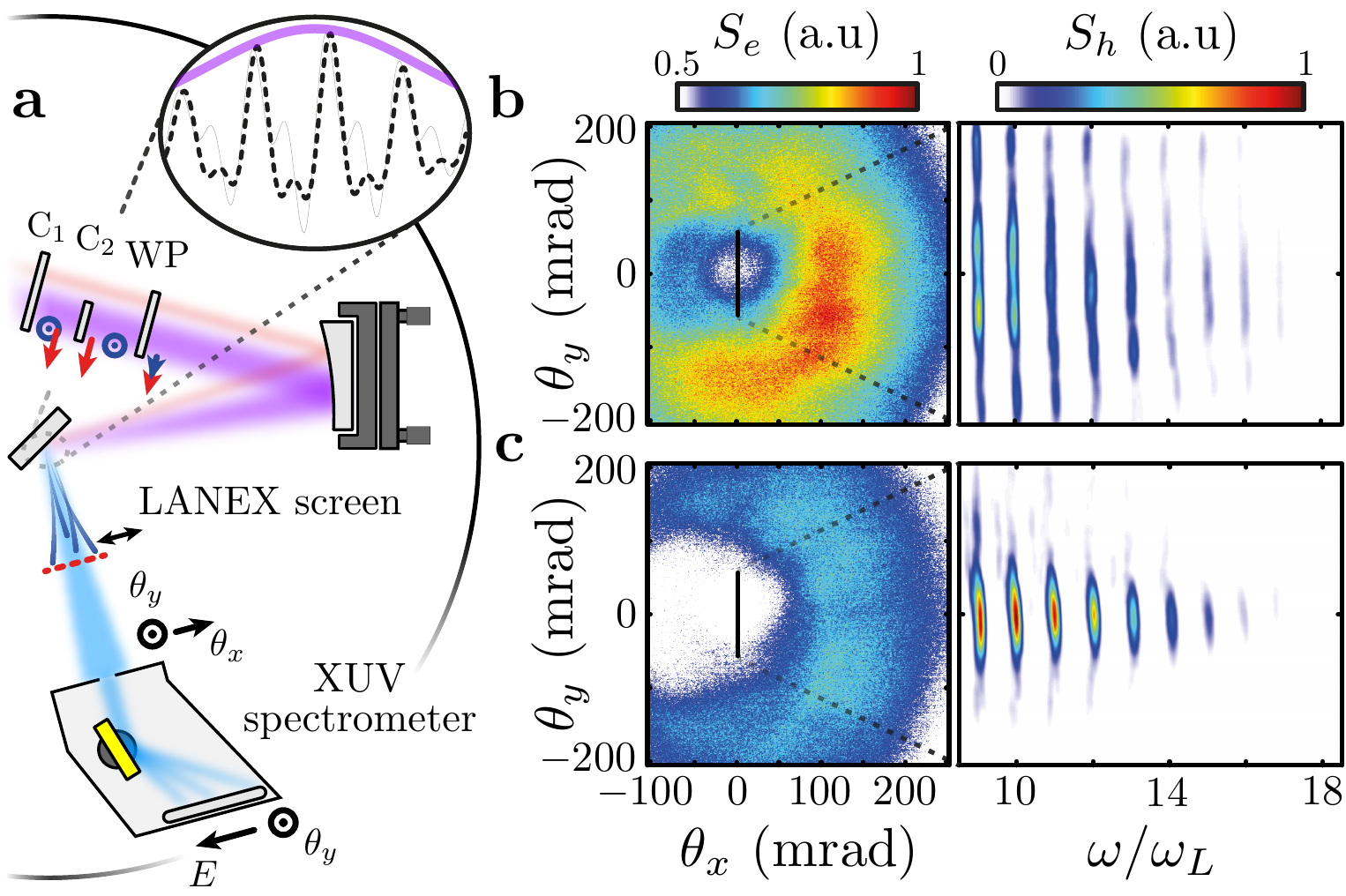}
	\caption{
	(\textbf{a}) Experimental setup with the prepulse (red) and main pulse (purple). The inset shows waveforms obtained for $\tau_{\phi,h} = 0$ (solid gray line) and $\tau_{\phi,e} = -T'/4$ (dashed black line).  Angular emission pattern of accelerated electrons (left) and angularly-resolved harmonic spectrum (right) obtained with a gradient scale length $L_g \lesssim \lambda_L/50$ and for a phase delay maximizing electron emission $\tau_{\phi,e}$, (\textbf{b}), or the high-order harmonic emission $\tau_{\phi,h}$ (\textbf{c}). The vertical black lines in the electron beam profiles mark the XUV spectrometer angular acceptance.
	}
	\label{fig:figure1}
\end{figure}

The two-color waveforms of total energy of $\approx 125\:$mJ are generated by a combination of a KDP frequency-doubling crystal (C$_1$), a calcite crystal (C$_2$) for timing control, and finally a quartz wave plate (WP) for setting parallel polarization directions of both color components (See SM~\cite{SM}).
A small $0.13^\circ$-rotation of the calcite plate about the laser polarization axis leads to a quasi-linear shift of the relative group and phase delay of the two color components by $\tau_\mathrm{g}\approx1.7\:$fs and $\tau_\phi=1.34\:$fs, respectively. 
The latter corresponds to one second-harmonic period, $T' = 2\pi/2\omega_L$. Such a rotation therefore scans the full range of two-color optical-cycle shapes with excellent temporal stability and reproducibility while keeping the pulse envelopes well overlapped.
Both the prepulse and the two-color main-pulse are p-polarized and focused by a $f = 200$ mm off-axis parabola onto a fused-silica target at $58^\circ$ angle of incidence. The measured diameter (FWHM in intensity) of the optimized focal spot of the fundamental color component is $5.6$ $\mu$m, whereas it is estimated to $3.3$ $\mu$m for the second harmonic beam, leading to on-target peak intensities of $5\times10^{18}$ and $3.5\times10^{18}$ W/cm$^2$, respectively. Two diagnostics for the plasma mirror emission have been implemented as displayed in Fig.\ref{fig:figure1}a. First, the spatial profile of the electron beam, $S_e(\theta_x,\theta_y)$, was measured using an insertable LANEX screen, placed at $150\:$mm from the target and imaged by a CCD camera. This scintillating screen is protected with a 13 $\mu$m thick aluminium foil and thus detects only electrons with energies $\gtrsim 0.15$ MeV. The high-harmonic emission, $S_{h}(\omega,\theta_y)$, was characterized using an angularly-resolved XUV spectrometer with an angular acceptance of 200~mrad.


As a first step, we used the main driving field of frequency $\omega_L$ only to measure the evolution of the experimental observables with the gradient scale length, in otherwise equal interaction conditions to the subsequent two-color experiments. Letting the prepulse arrive after the main pulse minimizes $L_g$ to $\lesssim \lambda_L/50$, limited by the main pulse temporal contrast. As demonstrated earlier \cite{kahaly2013direct, rodel2012harmonic}, such a steep gradient strongly favors harmonic generation by the \textit{Coherent Wake Emission} (CWE) process \cite{quere2006coherent}.
This mechanism leads to a sharp spectral cutoff at the maximum plasma frequency given by the target material density ($\sim 20\omega_L$ for silica), as well as rather high divergence of the emitted beam \cite{leblanc2017spatial}. For a smoother density gradient ($L_g > \lambda_L/50$), larger charge separation fields let the ROM process become predominant~\cite{rodel2012harmonic,kahaly2013direct,chopineau2019identification}, which at moderate intensities leads to a lower-divergence XUV beam \cite{dromey2009diffraction, vincenti2014optical}. In this experiment, we indeed observe this striking change of the XUV beam divergence (see SM~\cite{SM}). This evidences a transition from CWE to ROM harmonics with increasing $L_g$, although the spectra do not extend beyond the CWE spectral cutoff in our weakly relativistic regime ($a_0\approx1.5$). The correlated high-energy electron beam in the ROM-conditions ($L_g\approx\lambda/15$) was very similar to that measured in previous experiments~\cite{thevenet2016vacuum}, with a distinctive hole in the spatial beam profile centered on the specular direction, resulting from the interaction between the ejected electrons and the reflected laser light during VLA. Furthermore, the ejected charge and ROM harmonic efficiency are correlated insofar as both are maximized for $L_g \sim \lambda_L/15$, as observed before~\cite{chopineau2019identification}.


Using the temporally-shaped two-color main beam, this clear correlation disappears : while the ejected electron charge varies very similarly with $L_g$ as with the $\omega_L$-only driver, the harmonic signal drops with increasing $L_g$ to be finally absent around $L_g \sim \lambda_L/15$ (see SM~\cite{SM}). This is the case for all two-color phase delays $\tau_\phi$. The parameters optimizing the high-order harmonic generation are thus no longer the same as those for relativistic electron emission, which is a first indication that two-color drivers modifies the plasma mirror dynamics.


This is further corroborated by the measurement of the electron and harmonic signals for the shortest gradient ($L_g \lesssim \lambda_L/50$) as a function of the $\omega_L$--$2\omega_L$ phase delay $\tau_{\phi}$, i.e. of the optical cycle shape. The main experimental findings are shown in Fig.\ref{fig:figure1}b/c, by presenting the angular emission pattern of relativistic electrons $S_e(\theta_x,\theta_y)$ and the angularly-resolved harmonic spectrum $S_{h}(\omega,\theta_y)$ in the cases where $\tau_{\phi}$ is optimized for either the high-energy electrons or the high-order harmonics. In our interaction conditions, we find clearly distinct optima for both observables. In particular, when the optical-cycle shape maximizes the electron charge (Fig.\ref{fig:figure1}b), a highly divergent harmonic beam is reminiscent of that obtained in the CWE regime, whereas this beam becomes much less divergent and brighter in the other case (Fig.\ref{fig:figure1}c). This suggests that these harmonics are now generated via the ROM process, which had not been observed with the $\omega_L$-single-color driver for such a short gradient (see SM~\cite{SM}). We conclude that on top of a less-bright CWE background, the ROM harmonic generation shows a on-off oscillation as function of the driving optical cycle shape. No significant change in the electron spatial distribution other than the total charge is noticed. 


\begin{figure}
	\centering
		\includegraphics[width=1\columnwidth]{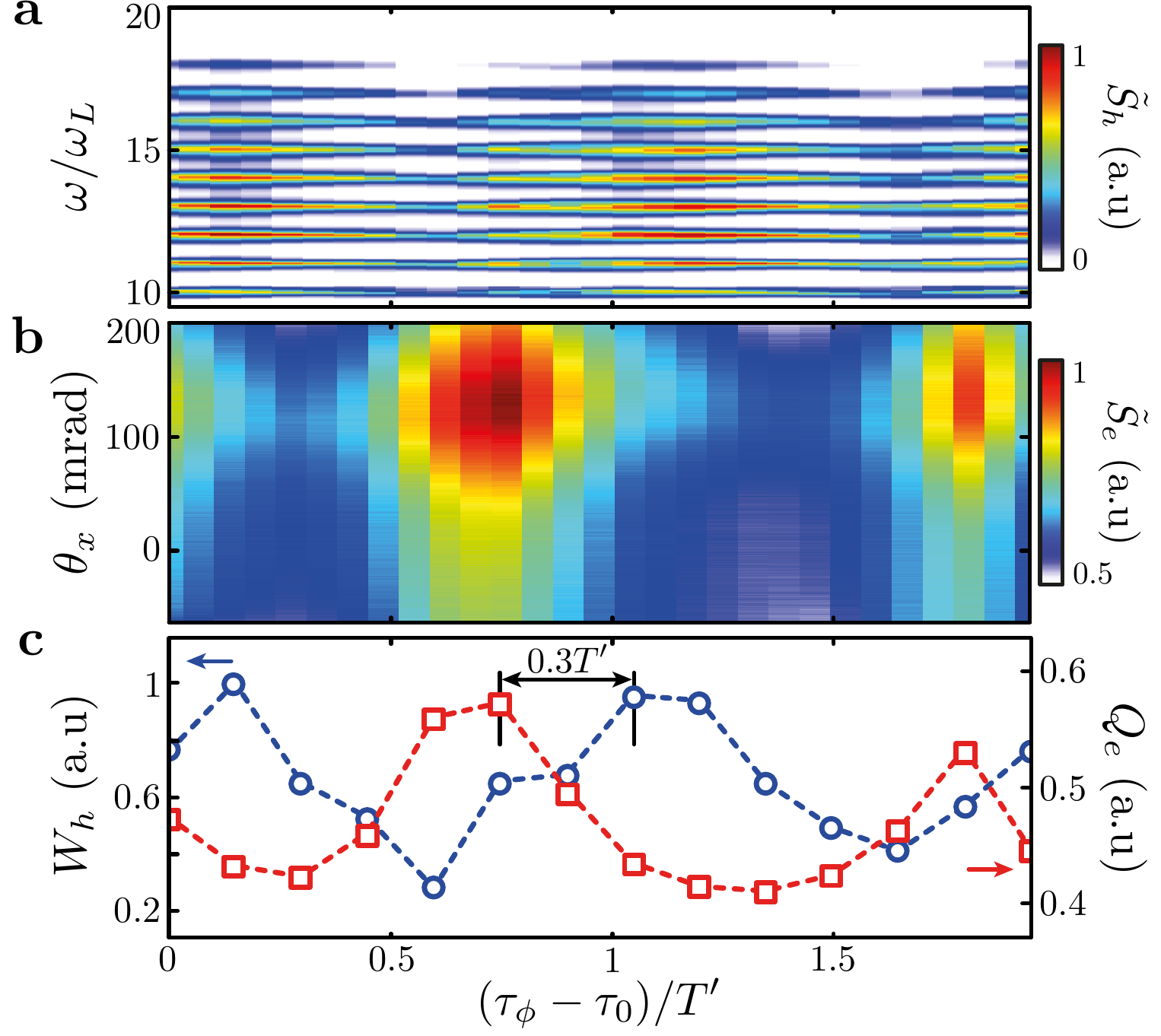}
	\caption{Experimental harmonic spectrum $\tilde{S}_{h}$ (\textbf{a}) and accelerated electron beam profile $\tilde{S}_e$ (\textbf{b}) in the incidence plane as a function of the phase delay $\tau_\phi$ (with an unknown experimental offset $\tau_0$). The total high-harmonic energy $W_{h}$ (blue circles) and electron charge $Q_e$ (red squares) obtained by integration over the vertical dimensions in the panels above are plotted in panel (\textbf{c}). The ejected electron charge and the harmonic energy similarly oscillate but with optimums shifted by $\simeq 0.3T'$. The experimental parameters are the same as in Fig.\ref{fig:figure1}.
	}
	\label{fig:figure2}
\end{figure}


The details of the experimental phase-delay dependence of the ejected electron charge and the harmonic emission are presented in Fig.\ref{fig:figure2}. First, the emitted harmonic spectrum $\tilde{S}_{h}(\tau_{\phi}, \omega)$, shown in Fig.\ref{fig:figure2}a, represents the spectral intensity within the central 33-mrad wide divergence cone, i.e. preferentially the ROM signal. The total high-harmonic energy $W_{h}(\tau_\phi)$ is obtained by integration over the presented spectral range ($\omega/\omega_L \in [10, 20]$). A clear modulation with a period $T' \approx 1.35\:$fs is observed (Fig.\ref{fig:figure2}c). 
The electron beam angular profile  $\tilde{S}_\mathrm{e}(\tau_\phi,\theta_\mathrm{x})$, obtained by integrating $S_e(\theta_x,\theta_y)$ over $\theta_\mathrm{y}$, is shown in Fig.\ref{fig:figure2}b. Further integration over $\theta_\mathrm{x}$ yieds the emitted electron charge $Q_{e}(\tau_\phi)$ shown in Fig.\ref{fig:figure2}c. It oscillates with the same period as the harmonic signal, with lower contrast but nonetheless presenting clear extrema, which had never been observed before. A striking finding is the dephasing of these periodic modulations: the optimum phase delays maximizing the ejected electron charge or ROM harmonic emission are shifted by $\approx 0.3T' = 0.4\:$fs. Therefore, intense attosecond light pulses are not necessarily generated in correlation with high-charge electron beams, in contrast to experiments where control is solely achieved through the plasma density gradient scale length $L_g$. Temporally shaping the driving optical cycles thus adds a new dimension to the control parameter space, which allows optimizing properties of the ejected electron bunches whose effect on HHG outweighs that of their total charge.


\begin{figure}
	\centering
		\includegraphics[width=1\columnwidth]{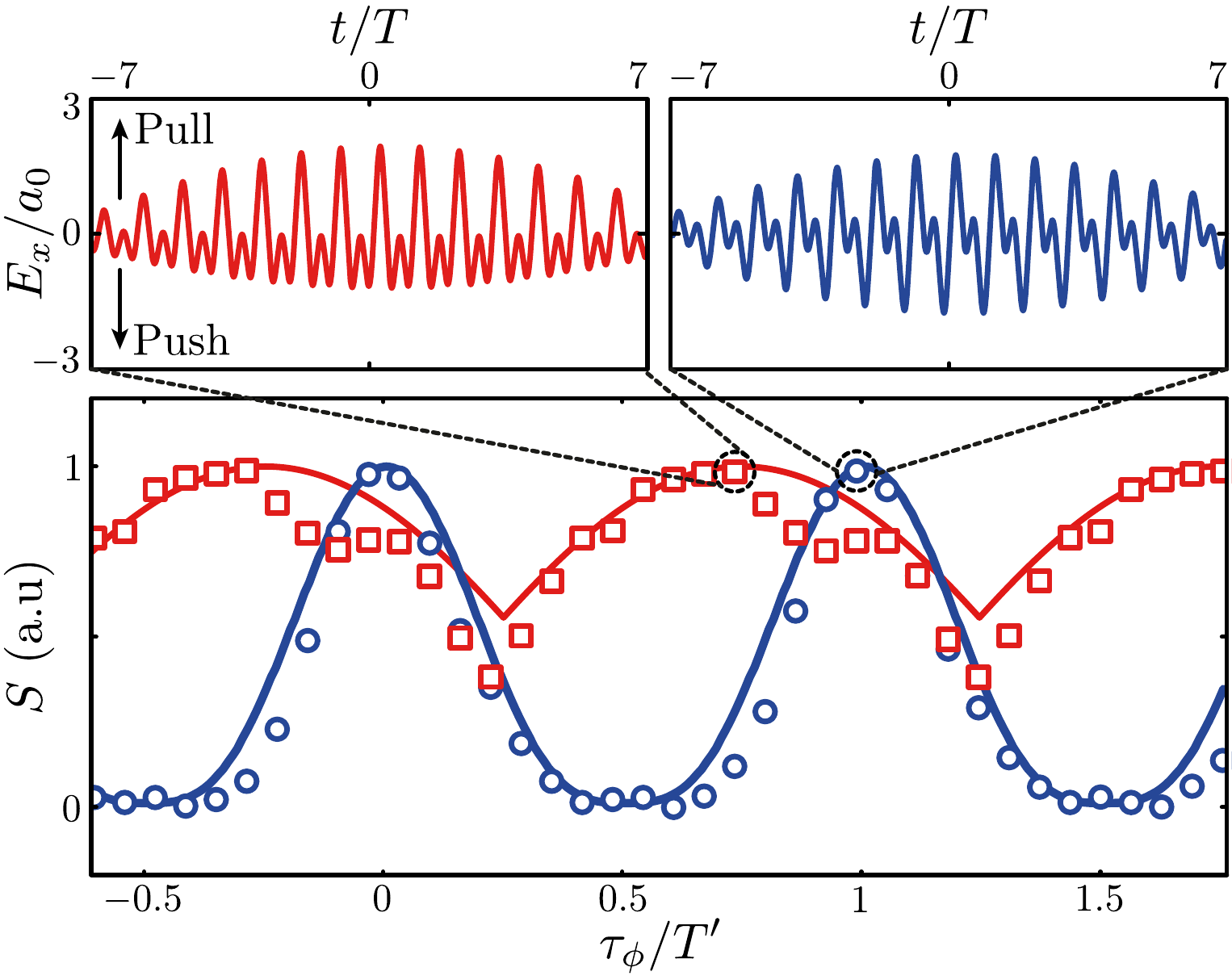}
	\caption{
	Ejected electron charge (red squares) and high-order harmonic energy (blue circles) as function of the phase delay $\tau_\phi$, obtained from 2D PIC simulations. Full blue and red lines show the normalized heuristic model predictions $S(\tau_\phi)$ and $P(\tau_\phi)$ for E-field steepness ($n=9$ here) and maximum pulling E-field, respectively. The periodic modulations of these electronic and harmonic signals are shifted by $\simeq T'/4$. The insets above show the driving fields for maximum electron (left) and high-harmonic (right) emission. A positive-valued electric field pulls electrons out of the plasma into vacuum.
	} 
	\label{fig:figure3}
\end{figure}


In order to gain insight into the dynamically controlled sub-cycle plasma dynamics, we turn to Particle-In-Cell simulations using the WARP+PXR code~\cite{warp, picsar, vay2016recent, vincenti2016detailed, blaclard2017pseudospectral, vincenti2017efficient, vincenti2018ultrahigh}. We consider a p-polarized incident laser pulse with electric field $ E(t, \tau_\phi) =  E_1(t)\sin(\omega_\mathrm{L}t) + E_2(t+ \tau_\phi) \sin[2\omega_\mathrm{L} ( t + \tau_\phi)]$, where $E_1(t)$ and $E_2(t)$ are 18-cycles long cosine-half-cycle envelopes ($\approx 30\:$fs FWHM) with equal peak field strengths corresponding to a normalized vector potential of $a_0=1$ for an 800-nm wave. It impinges a solid target with gradient scale length $L_g = \lambda_L/50$ at $55^\circ$ angle of incidence. The main results of 2D PIC simulations are summarized in Fig.\ref{fig:figure3}, which shows the ejected electron charge (only electrons with energies $\gtrsim$ 0.1 MeV), as well as the high-order harmonic energy (10th order only for clarity) as a function of the phase delay $\tau_\phi$. Similarly to the experiments (although with higher contrast), we retrieve clear oscillations of the harmonic and electron signals with period $T'$ and a relative shift by $\approx T'/4 = 0.35\:$fs, in good agreement with the experimental results. Also shown in Fig.~\ref{fig:figure3} are the incident driving waveforms that maximize the harmonic and electron emissions. In agreement with previous simulations done for similar as well as much higher laser intensities~\cite{yeung2017experimental,edwards2014enhanced,edwards2016waveform}, we find that the harmonic emission is optimized for a maximized waveform steepness in the change from the push to the pull phase (from negative to positive E-field). The emitted electron charge is found to be maximized, in our weakly relativistic conditions, rather for a waveform with the strongest possible field in the pull-phase. This suggests a simple heuristic model: \emph{(i)} a power-law dependence on the E-field steepness, $S(\tau_\phi)=\left(\mathrm{max}_t\left[\partial E(t, \tau_\phi)/\partial t\vert_{E(t)=0}\right]\right)^n$, describes the harmonic signal modulation, and \emph{(ii)} the maximum pulling field $P(\tau_\phi) = \mathrm{max}_t[E(t,\tau_\phi)]$, describes the electron charge modulation. Despite the extreme simplicity, these predictions are found to fit the signal modulations remarkably well, as shown in Fig.~\ref{fig:figure3}.


\begin{figure}[t]
	\centering
	\includegraphics[width=1\columnwidth]{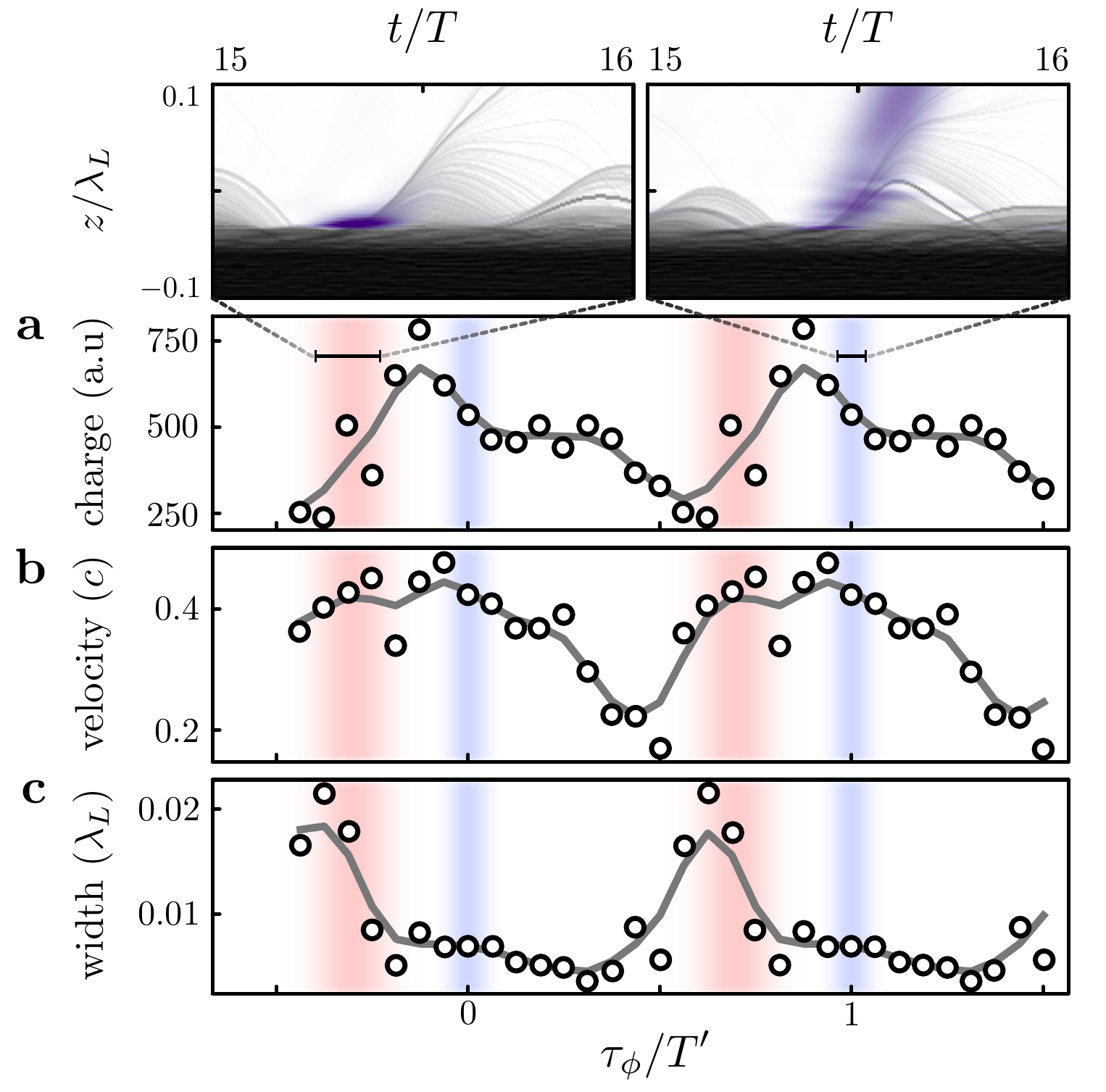}
	\caption{Electron bunch properties during the pull phase as a function of the phase delay $\tau_\phi$, extracted from 1D PIC simulations: mean charge (\textbf{a}), mean velocity (\textbf{b}), and initial spatial width (\textbf{c}). Blue and red shadings mark the phase delay ranges maximizing harmonic and electron emissions, respectively, in the 2D simulations as shown in Fig.\ref{fig:figure3}. The insets show the temporal evolution of the plasma electron density (gray scale color map, in log scale) in each phase delay range, spatially resolved along the normal to the target surface. The emitted attosecond pulses are overlaid to this density map in purple.}
	\label{fig:figure4}
\end{figure}


To gain a deeper physical insight, we have performed 1D PIC simulations which allow performing larger ensemble runs with better statistics. The simulated conditions are kept the same as in the 2D case discussed above. For each driving waveform characterized by $\tau_\phi$, we analyze the electron bunch properties during the central optical cycle. At every instant, only those electrons with velocity $>0.1c$ towards the vacuum are selected to calculate their total charge, the bunch position as their spatial center of mass, and the bunch width as their spatial root mean square width. The time derivative of the bunch position gives the bunch velocity. We then select a time interval $\Delta t_\mathrm{p}$ for the pull phase, starting at the instant of maximum bunch charge and ending at the instant when either the bunch velocity vanishes or when the bunch position has reached $z=0.25\lambda_L$. The bunch properties plotted in Fig.\ref{fig:figure4} as function of $\tau_\phi$ are the bunch charge and velocity, both averaged over $\Delta t_\mathrm{b}$, and the bunch width at the beginning of $\Delta t_\mathrm{b}$. Note that this bunch charge is not directly that of the emitted VLA electron beam: the bunch analyzed here has not yet left the plasma and, as clearly visible in the insets in Fig.\ref{fig:figure4}, part of it will turn back into the plasma during the subsequent pushing phase, in particular in our weakly relativistic conditions.

As in the 2D-simulations, the harmonic emission 
is maximized around $\tau_\phi=0$ and thus for the fastest switch from pushing to pulling field. While this does not maximize the electron bunch charge, it creates both a well compressed and fast outgoing electron bunch, which are the key parameters for efficient ROM HHG~\cite{edwards2014enhanced, edwards2016waveform}. The electron velocity alone, as considered in \cite{yeung2017experimental}, is not a sufficient criterion: while it is also boosted by the waveforms with enhanced pulling field, around $\tau_\phi=-T'/4$, this comes at the expense of a weakened push phase and thus a less compressed electron bunch. This is the situation we find to be optimal for electron emission in our conditions. As apparent in the left inset in Fig.\ref{fig:figure4}, the weaker field in the subsequent pushing phase lets a greater proportion of the bunch electrons fully escape the plasma with high velocity to be injected into the reflected laser field for VLA. Putting all emphasis on the pushing field strength ($\tau_\phi=+T'/4$), leads to the narrowest bunch, but at the expense of a smaller velocity. Finally, around $\tau_\phi=+T'/2$, where the harmonic emission is minimized, pushing and pulling peak fields are approximately balanced but the switch between the two is so slow, that the compressed electron bunch breaks apart and is not effectively accelerated outward. 


In conclusion, we used ultra-intense temporally shaped optical cycles to experimentally realize the dynamic control on the sub-femtosecond time scale of collective plasma electron motion on plasma mirrors emitting high-order harmonics and relativistic electrons. With the help of PIC simulations, we could elucidate how tuning the relative strength of pushing and pulling field half-cycle and the rapidity of the switch between them controls the formation and outward acceleration of the dense electron bunches that lead to HHG and seed VLA. These results illustrate the new possibilities opened by the attosecond steering of collective plasma electron motion driven by ultra-intense multi-color waveforms. We expect that expanding this approach to other frequency combinations and polarization shaping will have a strong impact on the development of powerful plasma-based radiation and particle sources.


\acknowledgements
We thank Aur\'elien Houard of LOA for lending the KDP crystal. This work was funded by the European Research Council under Contract No. 694596 of the European Union Horizon 2020 Research and Innovation Programme, the EquipEx Attolab (ANR-11-EQPX-005-ATTOLAB), the LabEx PALM (ANR-10-LABX-0039-PALM) and the R\'egion \^Ile-de-France (SESAME-2012- ATTOLITE). An award of computer time was provided by the INCITE programme (project PlasmInSilico). This research used resources of the Argonne Leadership Computing Facility, which is a DOE Office of Science User Facility supported under contract DE-AC02-06CH11357.


\bibliographystyle{h-physrev3}
\bibliography{Biblio}

\end{document}